\documentclass[journal]{IEEEtran}
\usepackage{graphicx}
\usepackage{dcolumn}
\usepackage{bm}
\usepackage{amssymb}
\usepackage{amsfonts}
\usepackage{amsmath}

\newcommand{\Tr}        {\mathrm{Tr}}

\newcommand{\ket}[1]    {\left| #1 \right\rangle}

\newcommand{\kb}[2]     {\left| #1 \right\rangle \!\! \left \langle #2 \right|}

\newcommand{\cM}        {{\mathcal M}}
\newcommand{\cN}        {{\mathcal N}}

\newtheorem{dfn}{Definition}

\begin{document}
\title{An Adaptive Entanglement Distillation Scheme Using Quantum Low Density
 Parity Check Codes}
\author{
\authorblockN{K. H. Ho and~H. F. Chau}\\
\authorblockA{Department of Physics and Center of Theoretical and 
Computational Physics, \\ The University of Hong Kong, 
Pokfulam Road, Hong Kong}}

\maketitle

\begin{abstract}
Quantum low density parity check (QLDPC) codes are useful primitives for
quantum information processing because they can be encoded and decoded 
efficiently. Besides, the error correcting capability of a few QLDPC 
codes exceeds the quantum Gilbert-Varshamov bound.~\cite{MMM04a}  Here, 
we report a numerical performance analysis of an
adaptive entanglement distillation scheme using QLDPC codes.  In particular,
we find that the expected yield of our adaptive distillation
scheme to combat depolarization errors exceed that of Leung and
Shor~\cite{LS07a, LS07b} whenever the error 
probability is less than about 0.07 or greater than about 0.28.
This finding illustrates the effectiveness of using QLDPC codes in
entanglement distillation.
\end{abstract}


\IEEEpeerreviewmaketitle
\section{Introduction}

Armed with quantum computers, Alice and Bob want to share copies of high
fidelity Bell state through an unknown noisy quantum channel.  One way to do so
is to compare their measured error syndromes of their shares of the quantum
particles using a pre-determined quantum error-correcting code (QECC) and then
to perform the necessary error recoveries.  Thanks to the quantum
Gilbert-Varshamov bound, there exists a QECC that produces perfect copies of
Bell state provided that the quantum error rate of the noisy channel is less
than about 19\%.  However, finding such a QECC as well as executing the
corresponding decoder can both be computationally intractable.  Another way to share copies of high
fidelity Bell state is to apply entanglement distillation purification (EDP) 
such as the recurrence method~\cite{BBP+96a,BDSW96a}. More precisely,
by two-way local operations and classical communications (LOCC2), Alice and Bob
discard those particles whose measurement results are not consistent with that
of the corresponding Bell states.  Thus, two-way EDP can be regarded as
a carefully designed quantum-error-detection-code-based error rejection method.
It can tolerate a higher error level than any one-way QECC-based method at the
expense of having a much lower yield.~\cite{BBP+96a,BDSW96a,M03}

QECC- and EDP-based entanglement distillation methods can be extended in many
ways. For instance, Gottesman and Lo~\cite{GL03a} as well as 
Chau~\cite{Chau02} introduced adaptive schemes using both QECCs and EDPs to
distill copies of almost prefect EPR pairs. Their schemes increase the error tolerance
level at the expense of the yield. Along a different line, Vollbrecht and 
Verstraete~\cite{VV05} as well as Hostens \emph{et al.}~\cite{HDM06} generalized
the recurrence method to raise the yield of entanglement distillation.
Recently, Leung and Shor~\cite{LS07b} introduced an entanglement distillation
protocol extending the earlier works of Maneva and Smolin~\cite{MS00} as 
well as Leung and Shor~\cite{LS07a}. Specifically, Leung and Shor~\cite{LS07b}
used a carefully constructed adaptive EDP-based protocol with universal hashing to
increase the yield over a certain range of channel error rates.

It is instructive to find a way for Alice and Bob to share Bell
states with an even higher yield without sacrificing their fidelity too much.
Naively, Alice and Bob may 
optimize the yield by estimating the noise level of the quantum channel before 
choosing the appropriate method.  However, this method is not ideal as the 
noise level of the quantum channel may change, say, in the presence 
of an adversary.

In view of the similarity between QECC-based and EDP-based schemes, it is
instructive to study methods that estimate the error rate and perform the
necessary error recovery or error rejection simultaneously.  Let us use the
following setting as the basis of our investigation.  Alice
and Bob pick a QECC.  They compare their error syndrome measurements and use
them to decide which qubits have to be discarded and which have to be
error-corrected.  However, not every QECC ${\mathcal C}$ is suitable for this
purpose because the error-correcting capability of a typical subcode of
${\mathcal C}$ formed by
puncturing a few discarded qubits may be drastically reduced.  Fortunately,
quantum low density parity check (QLDPC) code is ideal for this job.  First,
owing to the fact that QLDPC codes have sparse parity check matrices,
their average error-correcting capabilities do not in general change
greatly with the deletion of a few qubits.  Second, QLDPC codes can be
efficiently constructed~\cite{MMM04a,COT05a}. Third, efficient approximate
decoding methods such as belief propagation for classical low density parity
check codes (or LDPC codes for short)~\cite{P88, MM98, Mac99} can be readily
extended to QLDPC codes.  Therefore, it makes sense to investigate the
performance of a QLDPC-code-based entanglement distillation scheme. 
In fact, a preliminary study along this line by one group has been reported
in~\cite{HC07a}.

In Sec.~\ref{Sec:Review}, we briefly review the existing literature
of LDPC, QLDPC as well as the construction of the EDP from QECC.
In Sec.~\ref{Sec:Scheme}, we introduce a QLDPC-code-based entanglement
distillation scheme with LOCC2 as well as the performance
indicators in our analysis.  Then 
we study the performance of our scheme to combat depolarization errors
numerically in Sec.~\ref{Sec:Num}.  In particular, we
find that our scheme has a better yield than the recent method by Leung and 
Shor~\cite{LS07a, LS07b} to combat depolarization errors whenever the error
probability is less than about $0.07$ or greater than about $0.28$.
Finally, we conclude by giving the reasons why our scheme has a high yield
in Sec.~\ref{Sec:Conc}.  We also suggest some possible future works on
QLDPC-code-based adaptive EDP there.

\section{Preliminaries and Prior Arts}\label{Sec:Review}
\subsection{Classical low density parity codes}
\begin{dfn} 
A classical low density parity check (LDPC) code is a linear block code over
a finite field $GF(q)$ that has a sparse parity check matrix. In particular, 
a $(d_v, d_c)$-regular LDPC code has a sparse parity check matrix $H$ with
$d_v$ non-zero entries in each column and $d_c$ non-zero entries in each
row.~\cite{Mac99, Gal62, Gal63a, MN96}
\end{dfn}

\par\medskip
LDPC code can be represented by the so-called Tanner graph. Recall 
that a Tanner graph of a linear code $\mathcal{C}$ with a parity check 
matrix $H$ is a bipartite graph with vertex set $V = V_1 \cup V_2$. 
Each variable node in $V_1$ is associated with a bit of the 
code represented by a column of $H$; and each check node in $V_2$ is
associated with a generator of the code represented by a row of $H$. 
There is an edge linking $i \in V_2$ and $j \in V_1$ if and only if $H_{ij}
\neq 0$.

Let ${\mathcal C}$ be a LDPC code encoding $k$ bits of information as an
$n$-bit string by a sparse parity check matrix $H$. We denote
the encoded message by the column vector $\textrm{\boldmath $t$}$. After passing
this encoded message through a noisy channel, the receiver gets
the column vector $\textrm{\boldmath $r$} = \textrm{\boldmath $t$} +
\textrm{\boldmath $e$}$ where $\textrm{\boldmath $e$}$
is what we called the noise vector. The task of a decoder, therefore,
is to infer $\textrm{\boldmath $x$}$ given the error syndrome
$\textrm{\boldmath $s$} = H \textrm{\boldmath $r$}$ and the 
assumed properties of the channel. More precisely, the
decoder returns a column vector $\textrm{\boldmath $x$}$ that maximizes the
posterior conditional probability $\textrm{Pr}(\textrm{\boldmath $x$}|
\textrm{\boldmath $s$},H)$ of finding $\textrm{\boldmath $x$}$ given the syndrome
$\textrm{\boldmath $s$}$ and the parity check matrix $H$.

Many efficient approximate decoding strategies for LDPC codes can be regarded as
message passing algorithms executed on the corresponding Tanner graph. 
Message-passing decoding generally begins with variable nodes
sending messages to their neighboring check nodes. Decoding continues with 
rounds of messages sending back and forth between the check nodes and the 
variable nodes; and new messages are computed by each node as functions of messages
previously sent to them. The decoding algorithm terminates if a tentative decoding is
found. 

Famous for its linear runtime in the codeword size $n$,
belief propagation is one of the most commonly used message passing algorithm in which 
the messages passed between nodes in a Tanner graph are conditional 
probabilities~\cite{P88, MM98, Mac99}. More importantly, we shall show
in Sec.~\ref{Sec:Scheme} that belief propagation algorithm is applicable
to quantum stabilizer codes whose generators of the stabilizer is sparse.

\subsection{Quantum low density parity check codes}
\begin{dfn}
A quantum low density parity check (QLDPC) code is a quantum stabilizer
block error-correcting code over a finite field $GF(q)$ that has a sparse
parity check matrix. In particular, a $(d_v, d_c)$-regular QLDPC code has a
sparse parity check matrix $H$ with a constant column weight $d_v$ and a
constant row weight $d_c$.~\cite{MMM04a, COT05a}
\end{dfn}

\par\medskip
For example, the quantum error-detection code associated with each round of
entanglement distillation by the recurrence method and the Leung and Shor method are 
$(1,2)$- and $(2,4)$-regular QLDPC codes, respectively. (In some sense, these two
codes are atypical QLDPC codes as they compose of tensor products of block codes of sizes $2$ and $4$,
respectively.) Actually, a large number of QLDPC codes 
exist for a sufficiently large block code size $n$. Existing ways to construct them include: 

\subsubsection{Dual-containing CSS codes} \label{se:bicycle}
MacKay \emph{et al.}~\cite{MMM04a} constructed a few $GF(4)$ QLDPC codes 
from Calderbank-Shor-Steane (CSS) codes~\cite{CS96a, Ste96b} with certain constraints on the 
global structure of their parity check matrices. They include: 

\newcounter{code}
\begin{list}{(\roman{code})}{\usecounter{code} \setlength{\rightmargin}{\leftmargin}}
\item Bicycle codes: 
To construct a $[n,n-k]$ bicycle $GF(4)$ QLDPC CSS code with row weight $d_c$, 
MacKay \emph{et al.} first selected a
random $(n/2) \times (n/2)$ cyclic binary sparse-matrix $C_{\textrm{B}}$ with row weight $d_c/2$ and
defined a $(n/2) \times n$ matrix $H$ by
\begin{equation}\label{eq:H_bi}
H = [C_{\textrm{B}}, C_{\textrm{B}}^T],
\end{equation}
where $C_{\textrm{B}}^T$ is the transpose of $C_{\textrm{B}}$. Then, they deleted
some rows from $H$ to
obtain a new matrix $H_\textrm{B}$ with $k$ rows. One can easily check that $H_\textrm{B}$
is self-dual in the sense that $H_\textrm{B}H_\textrm{B}^T =0$.
As a result, $H_\textrm{B}$ can be used to construct a $[n,n-2k]$ binary CSS code. 
MacKay \emph{et al.} further showed that the performance of some bicycle codes
is better than the Gilbert-Varshamov rate for binary CSS codes~\cite{MMM04a}. 

\item Uicycle codes:
A unicycle $GF(4)$ code is constructed by making use of a perfect difference
set over an additive group. All pairs of rows of the cyclic binary matrix 
$C_{\textrm{U}}$ that make from the perfect difference set have an overlap of one.
To make the matrix self-dual, a new column of all ones is appended to
the matrix $C_{\textrm{U}}$. Thus, every pair of distinct rows of the 
resultant matrix $H_{\textrm{U}}$ have even overlapping. The self-dual
matrix $H_{\textrm{U}}$ can then be used to construct a CSS-type QLDPC 
code~\cite{MMM04a}.
\end{list}

\subsubsection{Group theoretical construction}

Instead of using CSS codes as the starting point, Camara \emph{et al.} 
constructed QLDPC codes by selecting low weight generators of the stabilizer
using certain group theoretical method~\cite{COT05a}. Numerical simulations
of the performance of a $(4, 8)$- and a $(6, 12)$-regular QLDPC
codes using their group theoretical method on the depolarizing
channel can be found in Ref.~\cite{COT05a}.
\subsubsection{Quantum quasi-cyclic LDPC codes}
Hagiwara and Imai invented a CSS-type construction of quantum quasi-cyclic LDPC code.
Their construction is based on algebraic combinatorics, and the performance
of their codes was analyzed in Ref.~\cite{HI07}. Recently, 
Hsieh \emph{et al.} proposed and investigated a new type of QLDPC codes 
from classical quasi-cyclic LDPC codes~\cite{HBD08a}.

\subsubsection{QLDPC codes constructed from finite geometries}
Aly proposed and analyzed the performance of a class of QLDPC whose parity check matrix are adapted to
be self-orthogonal with containing only one cycle of length four.~\cite{SAA07a}
\subsubsection{Asymmetric QLDPC codes}
Sarvepalli \emph{et al.} constructed a CSS-type of asymmetric QLDPC based
on BCH and finite geometry LDPC codes to take account the asymmetry for
the occurrence of bit flip and phase flip errors.~\cite{SRK08a}

\subsection{Entanglement distillation with two-way classical communications
by quantum error-correcting codes}

The general procedure of an adaptive LOCC2 stabilizer-code-based entanglement
distillation purification (EDP2) protocol can be described
below~\cite{GL03a, AG06}.
Alice prepares $n$ Bell states and
sends the second halves to Bob. Alice and Bob measure up to $(n-k)$ commuting generators 
of the stabilizer code one by one. After measuring each generator, they 
may throw away some of their shared quantum particles upon comparing their measurement
results. Then they compute the error 
syndrome. The $(i+1)$th generator used may depend on the results obtained
in the first $i$ measurements as long as this generator commutes with
all the previously measured 
operators. In the last step of the protocol, Alice and Bob perform local unitary 
transformation based on their earlier measurement results to distill the $k$ 
almost perfect Bell states.  Note that all one-way QECC-based entanglement purification
schemes together with the two-way recurrence method introduced by
Bennett \emph{et al.}~\cite{BBP+96a,BDSW96a} and its various
extensions~\cite{LS07a,VV05,HDM06,MS00,DNM03} can all be regarded as special
cases of this EDP2 protocol.

\section{An adaptive entanglement distillation protocol using quantum low density parity 
check codes} \label{Sec:Scheme}

\subsection{Dual-containing quantum low density parity check stabilizer codes}
\label{Subsec:QLDPC_construction}
Among the existing QLDPC code construction schemes in the literature, some can only build CSS
codes and some may not be efficient. So, in order to increase the error-tolerant capacity of our
practical adaptive entanglement distillation protocol, we have to find a simple and efficient way
to construct a large number of QLDPC stabilizer codes.  
Actually, our QLDPC code construction works for any $q$-ary code where $q = p^n$ 
is a prime power. We follow the notation of Ashikhmin and Knill~\cite{AK01}
by defining the unitary operators $X_a$ and $Z_b$ acting on a $q$-dimensional
Hilbert space by 
\begin{equation}
X_a: \ket{i} \longmapsto \ket{i+a}
\end{equation}
and
\begin{equation}
Z_b: \ket{i} \longmapsto \varpi_p^{\Tr(ib)}\ket{i}
\end{equation}
for all $a, b, i \in GF(q)$, where $\varpi_p$ is the $p$th root of unity and 
\begin{equation}
\Tr (i) = i + i^p + \cdots + i^{p^{n-1}} \in GF(p)
\end{equation}
is the absolute trace of $i  \in GF(q)$. Note that all arithmetic inside the
state ket and in the exponent of $\varpi_p$ is performed in the finite
field $GF(q)$. We also identify the unitary operator $X_a Z_b$ with 
$a + b \omega_{q^2} \in GF(q^2)$ where $\omega_{q^2}$ is a fixed primitive
element in $GF(q^2)$. Using this identification, we may abuse our language by 
saying, for example, that a qubit has experienced an error $a + b \omega_{q^2}$.

Our QLDPC stabilizer code construction is an extension of the bicycle
QLDPC CSS code construction by MacKay \emph{et al.}~\cite{MMM04a} reviewed in
Sec.~\ref{Sec:Review}. And it comes from a
simple but important observation concerning the matrix $H$ in Eq.~(\ref{eq:H_bi}). 
Suppose the elements of the matrix $C_B$ satisfy $(C_B)_{i, j} 
\equiv (C_B)_{ij} = \alpha_{i -j}$ for some $\alpha_{i - j} \in GF(q^2)$.
So, for $1 \leq i, i', j \leq n/2$,
\begin{equation}
(C_\textrm{B})^T_{i, i + i' -j} = (C_\textrm{B})_{i',j}
\end{equation}
and 
\begin{equation}
(C_\textrm{B})^T_{i', i + i' -j} = (C_\textrm{B})_{i,j}.
\end{equation}
Then, rows of the bicyclic matrix $H = [C_{\textrm{B}}, C_{\textrm{B}}^T]$ 
are mutually orthogonal to each other with respected to the skew-symmetric inner product
\begin{equation}\label{eq:in_prod}
(a + b \omega_{q^2} | c + d \omega_{q^2}) \equiv \Tr ( ad - bc) \in GF(p)
\end{equation}
for all $a, b, c ,d \in GF(q)$, irrespective of whether $C_B$ is sparse
or not. Since 
\begin{equation}
X_c Z_d X_a Z_b = \varpi_p^{(a+ b \omega_{q^2}| c + d \omega_{q^2})} X_a Z_b X_c Z_d,
\end{equation}
the rows of $H$ can be identified as the generators of the stabilizer of a $q$-ary
QECC~\cite{AK01, CRSS98a}; and so is $H_\textrm{B}$, the matrix obtained by deleting a 
few rows of $H$. 

In particular, by choosing $(\alpha_i)_{i=1}^{n/2}$ to be a sparse vector whose
elements are in $GF(q^2)$ so that $C_\textrm{B}$ is a sparse
$(n/2) \times (n/2)$ matrix, $H_\textrm{B}$ becomes the parity
check matrix of a $q$-ary QLDPC code. More importantly, the $GF(q^2)$
QLDPC code constructed in this way is not necessarily a CSS code. 

Interestingly, we may build a large number of regular QLDPC codes using this modified
bicycle construction. The trick is to pick the sparse vector 
$(\alpha_i)_{i=1}^{n/2}$ in such a way that
\begin{equation} \label{eq:con1}
|\{ i : \alpha_{n'i + j} \}| = u
\end{equation}
for all $j$ with the constraint $(n / 2)$ is divisible by $n'$. Then it is easy to check
that the parity check matrix $H$ constructed is $(n'u, 2n'u)$-regular.
And, by deleting the $(in' + j)$th row of $H$ for $i \in \mathbb{N}$, $j \in J$
where $J$ is a proper subset of $\{1, 2, \cdots, n'\}$, the resultant
parity check matrix $H_B$ corresponds to a $([n' - |J|]u, 2 n' u)$-regular
$q$-ray QLDPC code. For instance, let $q = 2, n = 12, n' = 3, (\alpha_i) = (1, \omega_4, \omega_4^2, 0, 0, 0)$
where $\omega_4$ is a primitive element in $GF(4)$, and $J = \{ 3 \}$. Then our construction
gives the $(2, 6)$-regular binary QLDPC stabilizer (but non-CSS) code
\begin{equation}
\left[
\begin{array}{cccccc|cccccc}
1 &\omega_4 &\omega_4^2 &0 &0 &0 & 1 & 0 & 0 & 0 &\omega_4^2 & \omega_4 \\
0 &1 &\omega_4 &\omega_4^2 &0 &0 & \omega_4 &1 & 0 & 0 & 0 &\omega_4^2 \\
0& 0 & 0 &1 &\omega_4 &\omega_4^2 &0 & \omega_4^2 &\omega_4 & 1 & 0 & 0 \\
\omega_4^2& 0 & 0 & 0 &1 &\omega_4 &0 & 0  & \omega_4^2 &\omega_4 & 1 & 0 
\end{array}
\right].
\end{equation}

It is easy to check that the (quantum) rate of the $(d_v, d_c)$-regular QLDPC code
constructed in this way is greater than or equal to $1 - d_v / d_c$, where the
equality holds if any only if the rows of $H$ are linearly independent over
$GF(q)$. In our subsequent study, we only consider those $H$'s with
full rank so that their rate is equal to $ 1 - d_v/ d_c$. Surely, this
extra constraint on the choice of $H$ is not very restrictive as our
construction is likely to give $H$ with full rank anyway.

Note that for a typical sparse vector $(\alpha_i)_{i = 1}^{n/2}$ satisfying
Eq.~(\ref{eq:con1}), the number $|\{i: \alpha_i = \beta \}| / (n/2)$ is about the same
for all $\beta \in GF(q^2)^*$. To summarize, we have succeeded in constructing a large number of regular
$q$-ary QLDPC codes. The construction is very efficient. Besides, their
regularity and almost equal probability of occurrence of non-zero elements in 
$(\alpha_i)_{i=1}^{n/2}$ make them reasonably effective to combat quantum
errors.

\subsection{Belief propagation algorithm for quantum stabilizer codes}
Belief propagation algorithm for classical LDPC codes can be extended to the
case of stabilizer code as follows. (See also Ref.~\cite{LP08} for a description
of a similar belief algorithm applied to graph states.)
A stabilizer code $\mathcal{C}$ associated with a parity check matrix $H$
can be represented by a Tanner graph with vertex set 
$V = V_1 \cup V_2$. Each variable node in
$V_1$ is associated with a qubit of the code represented by a column
of $H$; and each check node in $V_2$ is associated with a generator
of the code represented by a row of $H$. There is an edge linking $i \in V_2$
and $j \in V_1$ if only if $H_{ij} \neq 0$. 

By passing the messages
between the nodes, the task of the belief propagation decoding algorithm 
is to infer a tentative decoding $\tilde{\textrm{\boldmath$x$}}$.
That is to say, $\tilde{\textrm{\boldmath$x$}}$ is the most likely value of error 
experienced by the shared EPR pairs 
based on the measured error syndrome vector 
\begin{equation}
\textrm{\boldmath$s$} \equiv (s_i)_{i \in V_2} = (\sum_{j \in V_1} (H_{ij} | e_j))_{i \in V_2},
\end{equation}
where the check node $s_i \in GF(p)$ is the $i$th component of the syndrome
$\textrm{\boldmath$s$}$ and $e_j$ is the error experienced
by the variable node $x_j$. We call $\textrm{\boldmath $e$} \equiv (e_j)_{j \in V_1}$ 
the noise vector of the state shared by the sender and the receiver. 

The messages consist of two types of conditional probabilities $Q_{ij}^\alpha$
and $R_{ij}^\alpha$ associated with each non-zero entry in the parity check
matrix $H$ for all $\alpha \in GF(q^2)$. To aid discussions, we call
the $j$th component of the tentative decoding vector 
$\tilde{\textrm{\boldmath$x$}}$ the variable node
$\tilde{\textrm{\boldmath$x$}}_j \in GF(q^2)$.
The quality $Q_{ij}^\alpha$ approximates the belief that the qubit 
$\tilde{\textrm{\boldmath$x$}}_j$ has experienced the error  
$\alpha \in GF(q^2)$ given the messages received from all its checks other than $i$.
And the quality $R_{ij}^\alpha$ is the probability
of check $i$ being satisfied given that the variable node 
$\tilde{\textrm{\boldmath$x$}}_j$ has experienced an error in the state $\alpha \in GF(q^2)$ 
and the components of $\tilde{\textrm{\boldmath$x$}}$ other than
$\tilde{\textrm{\boldmath$x$}}_j$ have a separable distribution given by the \
probabilities $Q_{ij}^\alpha$'s. 

Initially, each message $Q_{ij}^\alpha$ is set to the prior probability $f_j^\alpha$ 
that $x_j$ has experienced an error $\alpha$. In situation of our interest, $f_j^\alpha$
is a quality of the quantum channel linking the two parties who would like to perform
entanglement distillation.
In each step, 
the quantities $R_{ij}^\alpha$ are updated according to the equation
\begin{equation}\label{eq:Rij}
R_{ij}^\alpha = \sum_{\textrm{\boldmath $x$}':x'_j=\alpha} \left[ {\rm Pr}(s_i|
\textrm{\boldmath $x$}')
\prod_{j' \in \cN(i)  \setminus\{j\} } Q_{ij'}^{x'_{j'}} \right] ,
\end{equation}
where
$\cN(i) \equiv \{ j:H_{ij} \neq 0 \}$ denotes the set of variable nodes
participated in the check $i$ and
\begin{equation}
\textrm{Pr}(s_i|\textrm{\boldmath $x$}') = \left \{
\begin{array}{ll}
1 & \mbox{if $\textrm{\boldmath $x$}'$ satisfies the check $i$}, \\
0 & \mbox{otherwise}.
\end{array}
\right.  
\end{equation}
That is to say,
\begin{eqnarray}
\textrm{Pr} (s_i | \textrm{\boldmath $x'$}) 
&=& \delta \left( \sum_{j' \in V_1} (
s_{ij'} | x'_{j'}), s_i \right)\nonumber \\
&=& \delta \left( \sum_{j' \in \cN(i) \setminus \{j\}} (
s_{ij'} | x'_{j'}), s_i - (s_{ij} | \alpha ) \right)
\end{eqnarray}
where 
\begin{equation}
\delta(x, y) = \left \{
\begin{array}{ll}
1 & \mbox{if $x = y$}, \\
0 & \mbox{otherwise},  
\end{array}
\right.  
\end{equation}
is the Kronecker delta.

For QLDPC stabilizer codes, Eq.~(\ref{eq:Rij}) 
can be computed efficiently using a fast-Fourier-transform-like recursive
iteration. In other words, we observe that 
\begin{equation}\label{eq:R}
R_{ij}^\alpha = R_{ij; \cN(i) \setminus \{j\}, s_i -(s_{ij}|\alpha)}
\end{equation}
where
\begin{equation}
R_{ij;J,b} = \sum_{ \{ x'_{j'} :  j' \in J \}}
\left[\delta\left( \sum_{j' \in J} \left(s_{ij'}|x'_{j'} \right),b \right) \prod_{j' \in J} Q_{ij'}^{x_{j'}'}\right]
\end{equation}
for all $b\in GF(p)$. Then we can evaluate Eq.~(\ref{eq:Rij}) by recursively applying the identity
\begin{equation}
R_{ij;J,b} = \sum_{c \in GF(p)} R_{ij;J_1,c}
R_{ij;J_2,b - c}
\end{equation}
for any partition $\{J_1, J_2 \}$ of the set $J$ with $|J_1| \approx |J_2|$
until $|J| = 1$. (And surely for $J = \{ j' \}$, $R_{ij;J,b}$ can be calculated directly
using Eq.~(\ref{eq:R}).)

After computing $R_{ij}^\alpha$ efficiently,
each check node $s_i$ sends the message $R_{ij}^\alpha$ to the variable node $x_j$ for all $j \in \cN(i)$.
Next, each variable node updates the messages
\begin{equation}\label{eq:Q1}
Q_{ij}^\alpha = \phi_{ij} f_j^\alpha \prod_{i' \in \mathcal{M}(j)\setminus\{i\}}
R_{i' j}^\alpha
\end{equation}
according to the information $R_{i' j}^\alpha$'s from check nodes $s_{i'}$'s for all
$i' \in \cM(j) \setminus \{ i\}$,
where $\cM(j) \equiv \{ i:H_{ij} \neq 0 \}$ is the set of checks involving variable
node $x_j$. The normalization constants $\phi_{ij}$'s ensure that the sum of 
conditional probabilities $\sum_{a \in GF(q^2)} Q_{ij}^\alpha = 1$. 

After each round of message passing, we compute the pseudo-posterior probabilities
\begin{equation}\label{eq:Q2}
Q_j^\alpha =  \phi_j f^\alpha\prod_{i \in \mathcal{M} (j)} R^\alpha_{ij},
\end{equation}
where $\phi_j$ is a normalization constant making $\sum_a Q_j^\alpha = 1$.
We now set $\tilde{x}_j$, the $j$th component of 
the tentative decoding $\tilde{\textrm{\boldmath $x$}}$, 
to $\alpha$ if $Q_j^\alpha \geq Q_j^\beta$ for all $b\in GF(q^2)$.  And we denote this
operation by
\begin{equation}\label{eq:argmax}
\tilde{x}_j = \raisebox{-1ex}
{\mbox{$\stackrel{\textrm{argmax}}{\scriptstyle \alpha \in GF(q^2)}$}}~ Q_j^\alpha.
\end{equation}

The decoding algorithm iterates until either the tentative decoding $\tilde{\textrm{\boldmath $x$}}$
is consistent with the observed syndrome (that is, 
$s_i = \sum_{j \in V_1}(H_{ij}|\tilde{x}_j ) $ for all $i \in V_2$) 
or a pre-determined maximum rounds of message passing is reached.

To summarize, the belief propagation algorithm can be
applied to decode QECC codes because its decisions depend only on our prior
assumptions of the noise of the channel and the measurement results for an
independent noise channel of the
error syndrome.  Moreover, it decodes QLDPC codes efficiently partly because
each summand in Eq.~(\ref{eq:Rij}) can be expressed as a sum of products.

\subsection{Our protocol}
After all the above preliminary discussions, we now report our adaptive
entanglement distillation scheme $\mathfrak{P}_\textrm{BP}$, which is an
EDP2 protocol using (binary) QLDPC codes to distill EPR pairs.

\par
\begin{figure}
\centering
\includegraphics*[scale = 0.3]{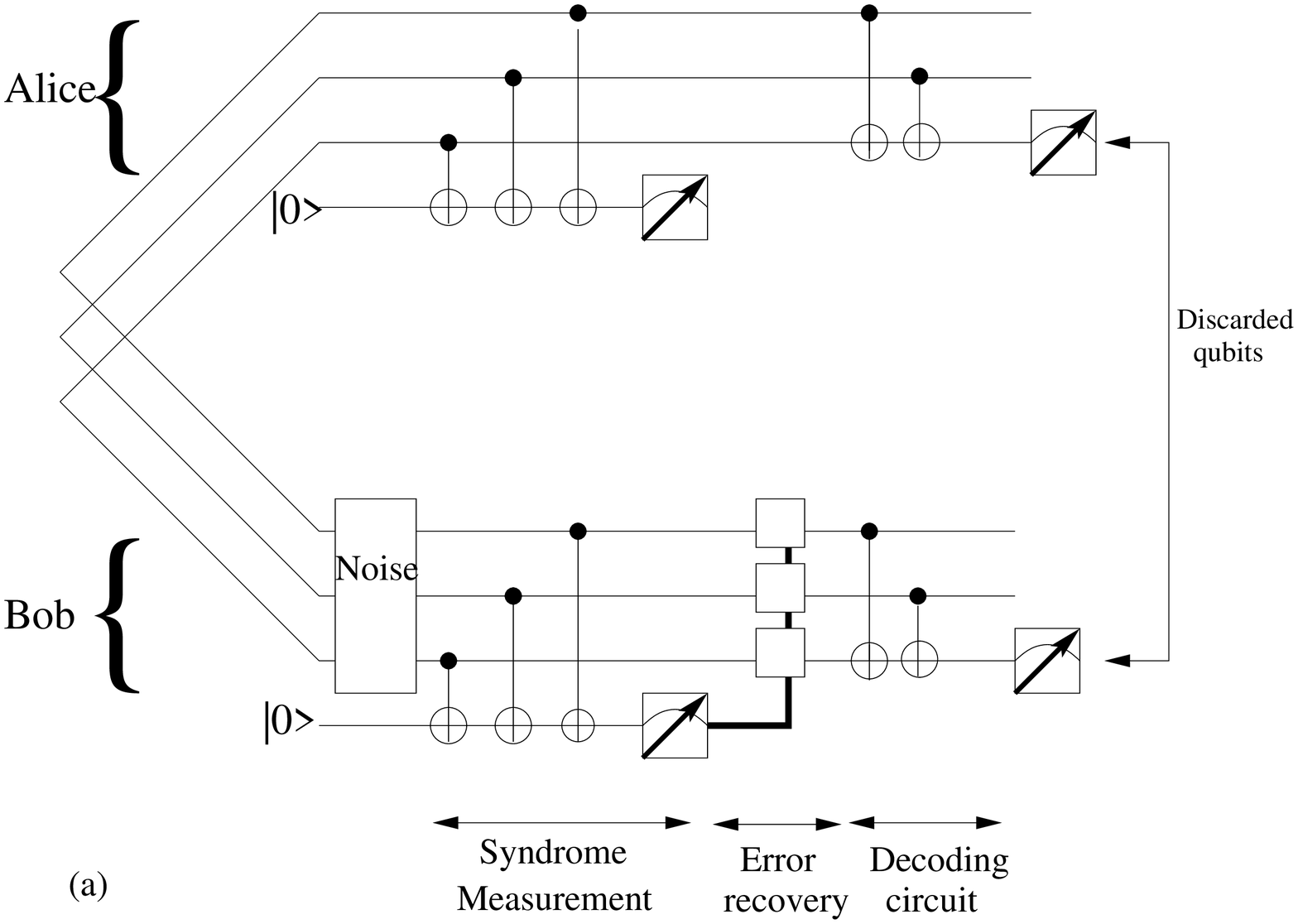}
\includegraphics*[scale = 0.3]{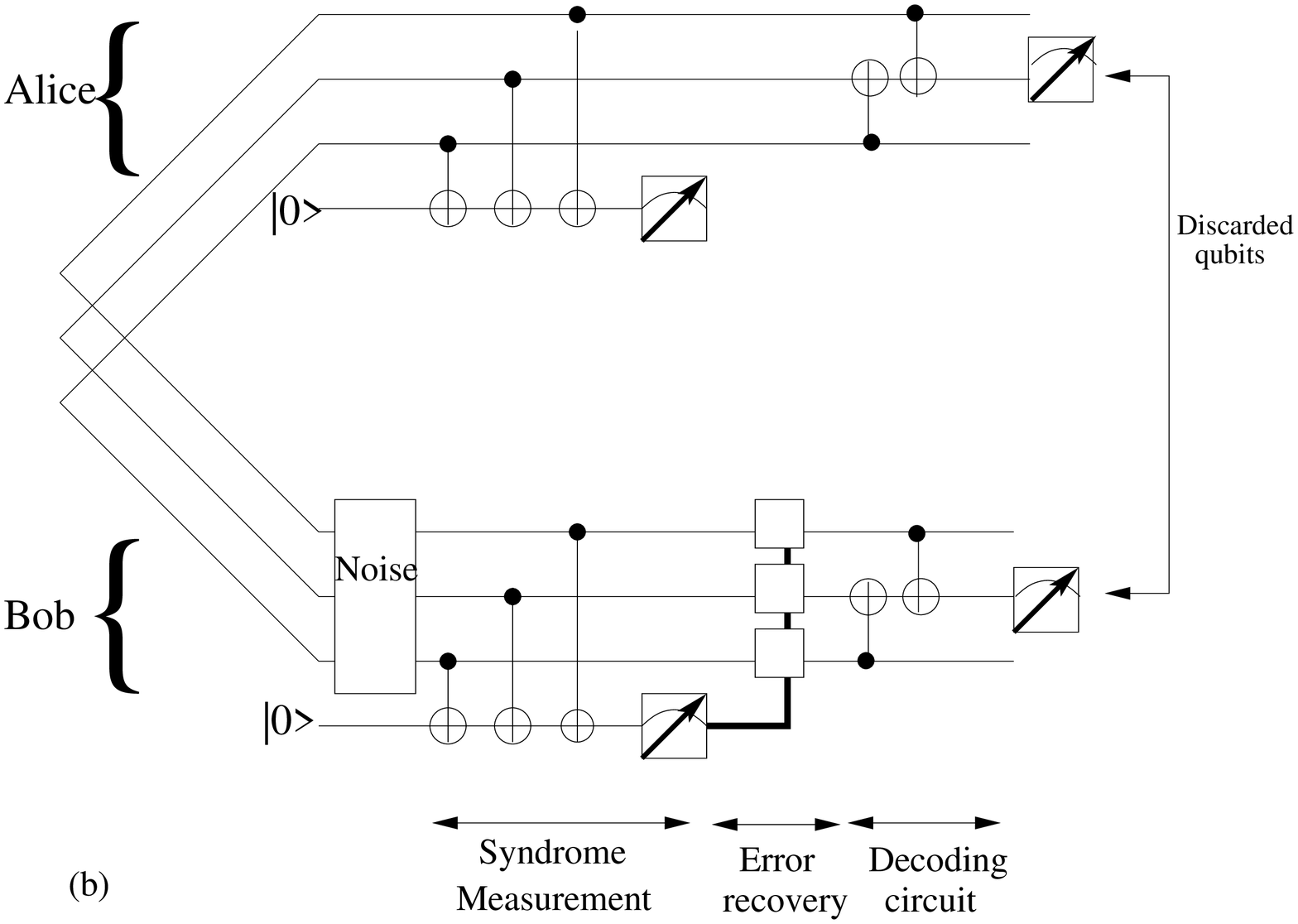}
 \caption{An illustrative example of ${\mathfrak P}_\textrm{BP}$ using $H[1] = 
(\omega_4 \; \omega_4 \; \omega_4)$. In the absence of noise, decoding circuits in (a) and (b) are
equivalent up to permutation of entangled qubits. But in the presence of noise, their performances
may differ due to the error propagation in the decoding process.}
\label{fig:de}
\end{figure}

\medskip
\begin{enumerate}
[{[The Adaptive Entanglement Distillation Scheme $\mathfrak{P}_\textrm{BP}$]}]
\item
 Alice prepares $n$ copies of EPR pairs $|\Psi^{+}\rangle \equiv \left(
 |00\rangle + |11\rangle \right) / \sqrt{2}$ and sends the second half of
 each pair to Bob through a noisy channel. Alice and Bob set the level $\ell$ to $1$.
\item \label{al:measure}
 Alice and Bob measure their corresponding shares of the noisy EPR pairs using 
 a pre-determined QLDPC code with a sparse parity check matrix $H[\ell]$ with the help of (unentangled) ancillas.
 Alice sends her measurement results to Bob.  And then Bob computes
 the error syndrome $\textrm{\boldmath$s$}[\ell]({\textrm{\boldmath $e$}})$, where ${\textrm{\boldmath $e$}}$ is the noise vector of
 the state they shared.
\item \label{al:bp}
 Using the belief propagation algorithm and Eq.~(\ref{eq:Q2}), Bob computes the posterior marginal
 probabilities $Q_j^\alpha [\ell]$ that his $j$th qubit has experienced an error
 $\alpha \in GF(4)$ given the messages passed from all its check nodes. From the posterior marginal probabilities, Bob deduces a tentative
 decoding $\tilde{\textrm{\boldmath$x$}} [\ell]$ based on the measured error syndrome
 $\textrm{\boldmath$s$}[\ell]({\textrm{\boldmath $e$}})$.
\item 
 If a tentative decoding $\tilde{\textrm{\boldmath$x$}} [\ell]$ satisfying
 $H[\ell] \tilde{\textrm{\boldmath$x$}}[\ell] = \textrm{\boldmath$s$}[\ell] ({\textrm{\boldmath $e$}})$ is found within the first
 $m_{\textrm{max}}$ rounds of message passing, then $\tilde{\textrm{\boldmath$x$}}[\ell]$
 is also a self-consistent error vector.  In this case,
 what Bob needs to do is to perform the error
 correction by applying the additive inverse of the pseudo-posterior noise
 vector, namely $-\tilde{\textrm{\boldmath$x$}} [\ell]$, to his qubits.  Finally, Alice and Bob finish
 up by running the encoding circuit for $H[\ell]$ backward to distill copies of almost
 perfect EPR pair. (See Fig.~\ref{fig:de}a.) This marks the end of our scheme.
\item \label{th:bp}
 If $H[\ell] \tilde{\textrm{\boldmath$x$}}[\ell] \neq \textrm{\boldmath$s$}[\ell] ({\textrm{\boldmath $e$}})$ even after
 $m_{\textrm{max}}$ rounds of belief propagation message passing,
 then Alice and Bob
 discard those EPR pairs whose believes of finding valid decodings are low.  More
 precisely, they fix an entropy threshold $h_\textrm{th}[\ell]$ and throw away
 the $j$th EPR
 pair if the entropy of the pseudo-posterior probabilities
 \begin{eqnarray}
  & & h_4(Q_j [\ell]) \nonumber \\
  & \equiv & h_4(\{ Q_j^\alpha [\ell] : \alpha \in GF(4) \} ) \nonumber \\
  & = & -\sum_{\alpha \in GF(4)} Q_j^\alpha [\ell] \log_2 Q_j^\alpha [\ell]
  \label{E:entropy_Q_j}
 \end{eqnarray}
 is greater than $h_\textrm{th} [\ell]$. 

The detailed procedure to throw away a EPR
 pair requires attention. According to the belief propagation algorithm, Alice and
 Bob believe that the most probable error experienced by the $j$th EPR pair is 
 $\alpha_j[\ell] = \{\alpha[\ell] \in GF(4): Q_j^\alpha[\ell] \geq Q_j^\beta[\ell], \forall
 \beta \in GF(4) \}$. So Bob first apply $-\alpha_j[\ell]$ to his share of the $j$th EPR
 pair. Surely, there are more than one possible encoding circuit for $H[\ell]$ and
 running any of these encoding circuit backward can correctly decode $H[\ell]$ in
 the absence of noise. Since the tentative decoding cannot be found, in order to
 minimize the decoding error, Alice and Bob run the encoding circuit backward in which the
 sum of the entropies of the pseudo-posterior probabilities for the message qubits are
 minimized. After applying this decoding circuit, they can throw away those shared EPR
 pairs with high entropy of the pseudo-posterior probabilities. (See Fig.~\ref{fig:de}.)
\item
 Alice and Bob increase the level $\ell$ by $1$.  If it exceeds a
 pre-determined number
 $\ell_{\textrm{max}}$, they give up all their shared particles
 and start over again.  Otherwise, they construct another sparse parity check matrix
 $H[\ell]$ orthogonal to $H[1], H[2], \ldots , H[\ell - 1]$ with
 respected to the skew-symmetric inner product in Eq.~(\ref{eq:in_prod}).  In general, the
 choice of $H[\ell]$ may depend on the marginal posterior probabilities
 $Q_j^\alpha [1]$'s, $Q_j^\alpha [2]$'s, $\ldots$ , $Q_j^\alpha [\ell - 1]$'s.
 They continue the decoding by going back to step~\ref{al:measure}.
\end{enumerate}

\par\medskip
Three remarks are in placed.  
First, for a sufficiently low channel error rate, a self-consistent
error vector is likely to be found without throwing away any EPR pair in
step~\ref{th:bp}.
Besides, this self-consistent vector is equal to the noise vector $\textrm{\boldmath$e$}$.
Consequently, our protocol ${\mathfrak P}_\textrm{BP}$ is reduced to a QECC-based scheme.  While for a sufficiently high
channel error rate, a self-consistent error vector is unlikely to be found.
Together with suitable choices of the entropy thresholds
$h_\textrm{th} [\ell]$'s, our protocol ${\mathfrak P}_\textrm{BP}$ becomes,
in effect, an EDP2 based scheme. In this respect, a 
row of $H[\ell]$ may be used either for error recovery or error rejection depending on the
error syndrome measurement results. This fulfills our goal of finding an adaptive 
entanglement distillation scheme that estimates the error rate and
performs the necessary error
recovery or error rejection simultaneously.  Second, our scheme can be
generalized to distilling generalized Bell states readily.
Third, $H[\ell]$'s should be picked in such a way that their error correcting
capabilities increase as the number of levels $\ell$ increases; and we report
a simple adaptive way to do so efficiently in the coming subsection.

\subsection{The choice of parameters for ${\mathfrak P}_\textrm{BP}$
 and the performance indicators}
\label{Subsec:Performance_indicator}

In this pilot study, we only consider the performance of our scheme for a
depolarizing channel, namely, each EPR pair has an equal and independent chance
of experiencing a Pauli error.  We denote the probability that a
qubit experience any one of the Pauli errors by $p_0$. Moreover, 
we do not focus on the performance of a particular QLDPC code used in our
EDP2 protocol.  Instead, we consider the average performance over an ensemble
of QLDPC codes used.  Moreover, these codes are randomly selected using
the method reported in Sec.~\ref{Subsec:QLDPC_construction}.

\subsubsection{The choice of parameters for ${\mathfrak P}_\textrm{BP}$}
The scheme ${\mathfrak P}_\textrm{BP}$ involves the use of QLDPC codes and an
adaptive procedure according to the error syndrome measurement results. We
try to understand the effects of these two ingredients on the performance
by studying two implementations of
${\mathfrak P}_\textrm{BP}$.

\medskip\par\noindent
{\it Implementation~A}:
\newcounter{par}
\begin{list}{(\roman{par})}{\usecounter{par} \setlength{\rightmargin}{\leftmargin}}
\item We fix the maximum number of levels $\ell_\textrm{max} = 1$.
\item We choose the QLDPC code $H[1]$ using the extended bicycle
construction reported in Sec.~\ref{Subsec:QLDPC_construction}.
\item \label{Sec:threshold} We set the entropy threshold $h_\textrm{th}[1]$
to
\begin{eqnarray}
h_\textrm{th} & \equiv & h_\textrm{th} [1] = S(W_{p_0}) \nonumber \\
& = & - (1 - p_0) \log ( 1- p_0) - p_0 \log \frac{p_0}{3} ,
\end{eqnarray}
where
\begin{eqnarray}
W_{p_0} &=& \frac{1 - p_0}{3} \kb{\Psi^+}{\Psi^+} + \frac{p_0}{3} \left( \kb{\Psi^-}{\Psi^-} \right.
+ \nonumber \\ & & \left.  \kb{\Phi^+}{\Phi^+} + \kb{\Phi^-}{\Phi^-} \right)
\end{eqnarray}
is the Werner state.
The rationale behind this choice is that
after passing through the depolarizing channel with quantum error
rate $p_0$, the density matrix $\kb{\Psi^+}{\Psi^+}$ becomes the Werner state
$W_{p_0}$. Therefore, in the absence of any additional information on the
errors syndrome measurement,
the entropy of the uncertainty of the kind of error experienced by each
$|\Psi^+\rangle$ is equal to $W_{p_0}$.
Thus, this choice of entropy threshold means that a pair is discarded only if
a self-consistent tentative decoding
$\tilde{\textrm{\boldmath$x$}} [1]$ cannot be found and the belief
propagation algorithm is unable to improve Bob's
knowledge on the kind of error the pair has experienced.
\item We put the maximum number of rounds of message passing $m_\textrm{max}
\approx 10$.  In fact, further increasing
$m_{\textrm{max}}$ does not improve the performance of message passing
algorithm to correctly find out the noise vector.
(See, for example, Ref.~\cite{Mac03a}).
\end{list}

\medskip
Clearly, Implementation~A can be used to study the performance of EDP2 using
QLDPC codes without an adaptive procedure. To study the power of adaptation, we
consider Implementation~B below.
\medskip\par\noindent
{\it Implementation~B:}
\setcounter{par}{0}
\begin{list}{(\roman{par})}{\usecounter{par} \setlength{\rightmargin}{\leftmargin}}
\item We fix $\ell_\textrm{max} = n/4$ where $n$ is
the codeword size of $H [\ell]$'s.
\item We first pick a QLDPC code $H$ using the
extended bicycle construction. From this $H$, we build a
hierarchy of codes as follow.
The first layer contains one code, namely, $H$ itself.  The $u$th
layer of codes are those formed by deleting exactly $u$ rows from $H$.
Moreover, a layer $u$ code $H_u$ is said to be connected to a layer
$(u + 1)$ code $H_{u+1}$ if $H_{u + 1}$ can be formed by removing
one row from $H_u$.
Now, we randomly pick a layer $\ell_\textrm{max}$ code in this
hierarchy to be our $H[1]$, namely, the QLDPC code used in the first level of
decoding in ${\mathfrak P}_\textrm{BP}$.
If the tentative decoding is not matched,
that is, $H[1] \tilde{\textrm{\boldmath$x$}}[1]
\neq \textrm{\boldmath$s$}[1] ({\textrm{\boldmath $e$}})$, then
the code $H[2]$ used in the second level of decoding in
${\mathfrak P}_\textrm{BP}$ is selected among those layer
$(\ell_\textrm{max} - 1)$ codes in the hierarchy that are connected to $H[1]$.
Surely, such a choice should, as far as possible, maximize the belief on the
errors experienced by the EPR pairs after running the belief propagation
algorithm with the code $H[2]$.
More precisely, $H[2]$ is chosen by adding one row of the parity check matrix
$H$ that is not present in $H[1]$.  And this additional row is selected so as
to maximize $\sum_{j\in V_2} h_4 (Q_j[\ell = 1])$, where $Q_j[\ell = 1]$ is the
pseudo-posterior probability obtained by running the belief propagation
algorithm with the code $H[1]$.  The codes $H[3], \ldots ,  H[\ell_\textrm{max}]$
are picked in a similar manner until either a consistent tentative decoding is found
or when $\ell = \ell_\textrm{max}$. In this way, we construct a sequence
of mutually orthogonal QLDPC codes adaptively and effectively.
\item We set $h_\textrm{th} [\ell_\textrm{max}] = S(W_{p_0})$.  More
importantly, we put $h_\textrm{th} [\ell] = 2$ for all $\ell <
\ell_\textrm{max}$, namely, the maximum possible value for a noisy EPR pair.
In this way, we avoid throwing away noisy EPR pairs prematurely.

\item As in Implementation~A, we set $m_\textrm{max} \approx 10$.
\end{list}

\medskip
The choice of $\ell_\textrm{max}$ in Implementation~B requires
clarification. In order to fully utilize the adaptive nature of
${\mathfrak P}_\textrm{BP}$, $\ell_\textrm{max}$ should not be too small.
Nevertheless, the average number of checks per variable node for $H[1]$ will
be less than $1$ if $\ell_\textrm{max}$ is about $n/2$, making $H[1]$
useless for distillation. In what follows, we take the middle path by fixing
$\ell_\textrm{max} = n/4$.

\subsubsection{Performance indicator}

The yield $D$ is used as the performance indicator. It is defined as the expected number of input
pairs needed per output perfect EPR pair in the limit of a large number of
input pairs.
The yield is high if the rate of the quantum code used is high and the
decoder error rate, namely, the chance for a qubit to be incorrectly decoded
is low.
We use the symbols $D_\textrm{A}$ and $D_\textrm{B}$ to denote the average
yields of our EDP2 protocol ${\mathfrak P}_\textrm{BP}$ using Implementations~A
and~B over the ensemble of QLDPC codes, respectively.

\section{Performance Of ${\mathfrak P}_\textrm{BP}$}
\label{Sec:Num}
We study the performance of ${\mathfrak P}_\textrm{BP}$ 
by numerical simulations.
Actually, our simulations show that the yields
$D_\textrm{A}$ and $D_\textrm{B}$ depend chiefly on the values of $d_v$ and
$d_c$ for the codes $H[1]$ and $H$ in Implementations~A and~B, respectively.
In other words, the yields are not sensitive to the actual sparse vector
$(\alpha_i)$ used in the extended bicycle code construction.
In all the figures below,
each data point represents the average yield $D_\textrm{BP}$
for $\mathfrak{P}_\textrm{BP}$ over $1000$ independently generated
noise vectors.  The associated one-sigma-level error bar is also shown.  

\begin{figure*}
\centering
\includegraphics*[bb = 15 30 460 400, scale = 0.8]{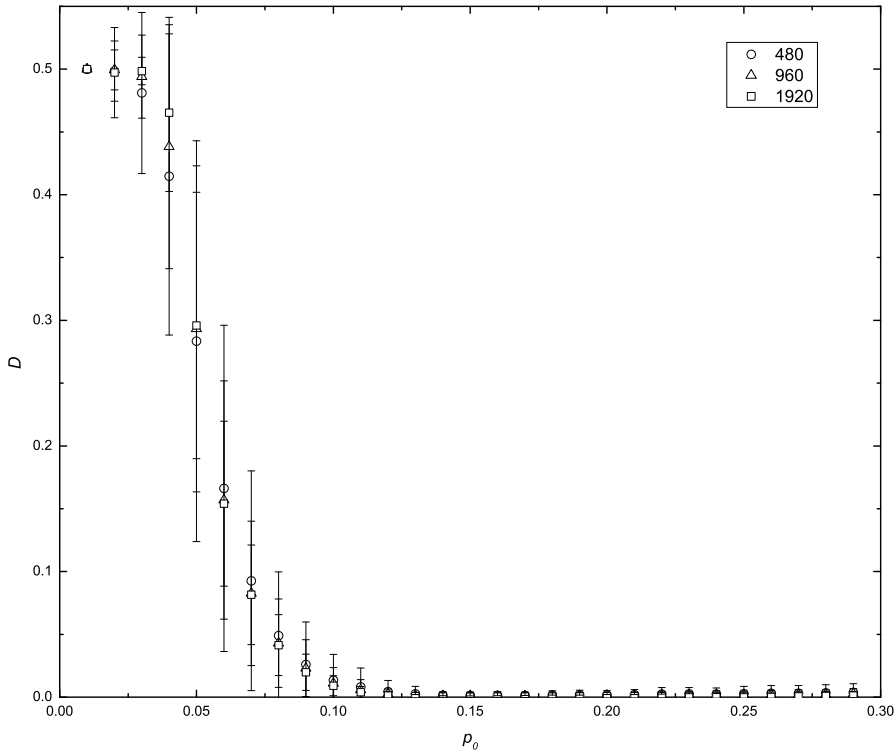}
 \caption{A plot of the yield $D_\textrm{A}$ of Implementation~A 
  using $(8, 16)$-regular QLDPC codes for different codeword sizes $n$ 
  against the error probability $p_0$.} 
\label{fig:ImA_n}
\end{figure*}

Our simulations show that, within $\approx 0.1\sigma$ level of uncertainty, the
yield $D$ of Implementations~A and~B do not depend on the codeword size $n$
provided that $n\gg d_v, d_c$. (See Fig.~\ref{fig:ImA_n}.) So, we fix the value
of $n = 960$ in all our subsequent discussions.

Another general feature concerning the error bars requires explanation.
As shown in Fig.~\ref{fig:ImA_n}, the sizes of error bars in a typical $D$
against $p_0$ plot depend strongly on the value of $p_0$.  When $p_0$ is
sufficiently small, the tentative decoding obtained from
${\mathfrak P}_\textrm{BP}$ correctly predicts the
errors experienced by the EPR pairs most of the time. Hence, the error
bar size is small.  When $p_0$ is so high that it cannot be handled by the
QLDPC code used, most of the EPR pairs will be thrown away after running
${\mathfrak P}_\textrm{BP}$. This makes both the yield $D$ and the
size of its error bar small. Interestingly, when $p_0$ is in between these two
extremes, the probabilities of correctly and incorrectly finding the
tentative decoding are comparable. More importantly, these two cases have
drastically different yields. As a result, the error bar of the average yield
$D$ in this regime is very large. In other words, the large variance of $D$ in this regime
is intrinsic and is not the result of insufficient sampling.

\begin{figure*}
\centering
\includegraphics*[bb = 10 10 461 395]{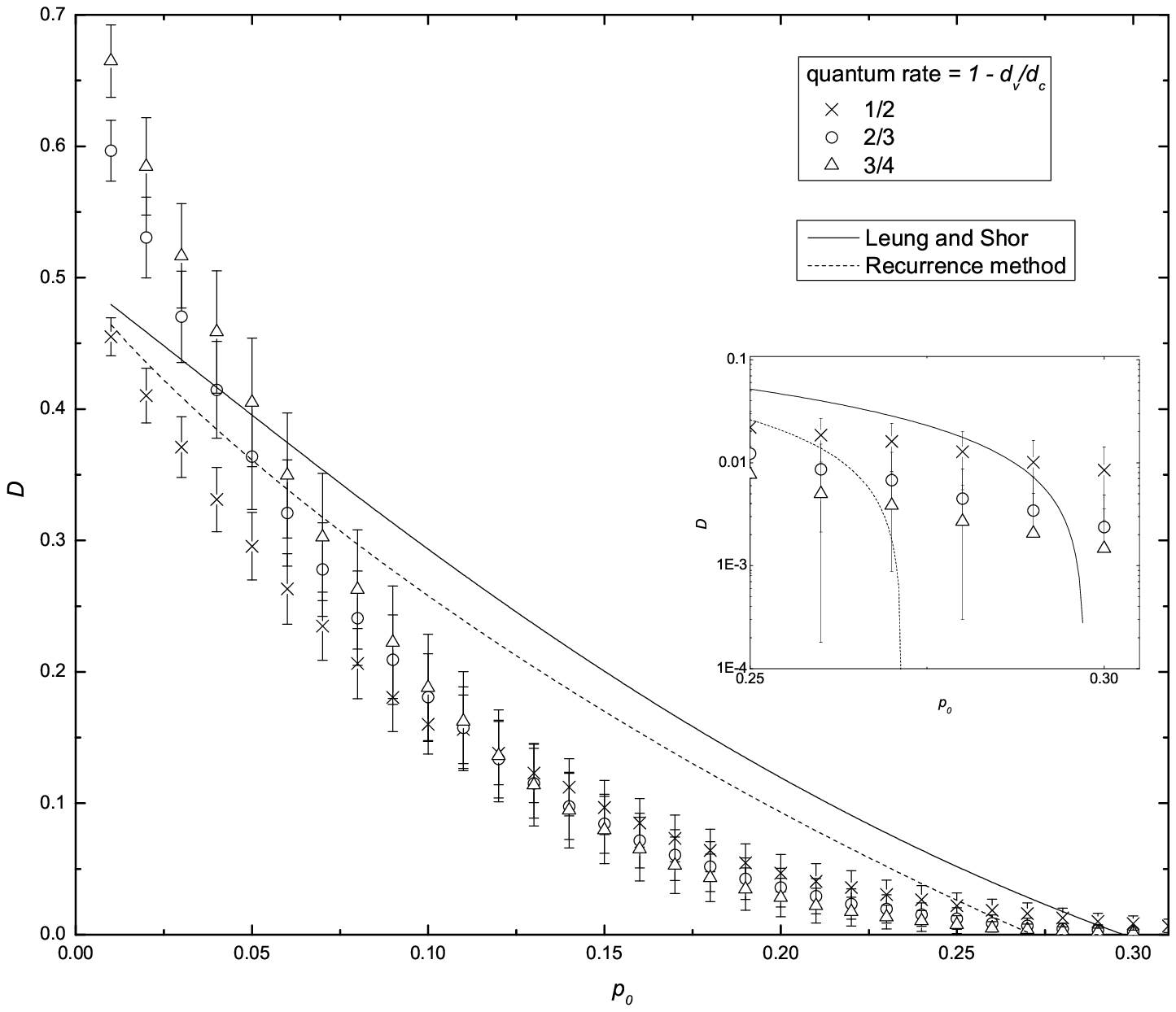}
 \caption{A plot of the yield $D_\textrm{A}$ of Implementation~A using
 $(d_v,d_c)$-regular QLDPC codes for different quantum rates against
 the error probability $p_0$. The values of $d_v$ and $n$ are fixed to be
 $2$ and $960$, respectively.}
\label{fig:ImA_dv2}
\end{figure*}

Let us compare the performances of Implementation~A using $(2,d_c)$-regular
QLDPC codes with the recurrence method~\cite{BDSW96a} and
its extension by Leung and Shor~\cite{LS07a}. This comparison makes sense
because of two reasons. First, the recurrence method makes use of
$(1,2)$-regular quantum codes with block size 2; and the Leung and Shor method
makes use of $(2,4)$-regular ones with block size 4.
Besides, all these three methods use quantum codes with minimum distance 2.
In other words, these codes can only detect but cannot correct quantum errors.
As shown in Fig.~\ref{fig:ImA_dv2}, the maximum error tolerable rate for
Implementation~A using $(2,4)$-regular QLDPC codes is higher than both the
recurrence and the Leung and Shor methods provided that the channel error rate
$p_0 > 0.28$. (In fact, our scheme can tolerate up to at least $p_0 = 0.30$.)
This result demonstrates the power of using
QLDPC codes to distill very noisy EPR pairs using two-way classical
communications. Fig.~\ref{fig:ImA_dv2} also depicts that by using
$(2,d_c)$-regular QLDPC codes with $d_c \geq 4$, the yield of Implementation~A is
higher than the recurrence method as well as the Leung and Shor's whenever
$p_0 < 0.05$.  This is not surprising because more EPR pairs are sampled
in each error syndrome measurement as $d_c$ increases. Nonetheless, it also
makes the yield decrease for a large value of $p_0$ because
propagation of quantum errors due to decoding in step~\ref{th:bp} of 
${\mathfrak P}_\textrm{BP}$
is more serious for a large $d_c$.

\begin{figure*}
\centering
\includegraphics*[bb = 20 25 490 662, scale = 0.8]{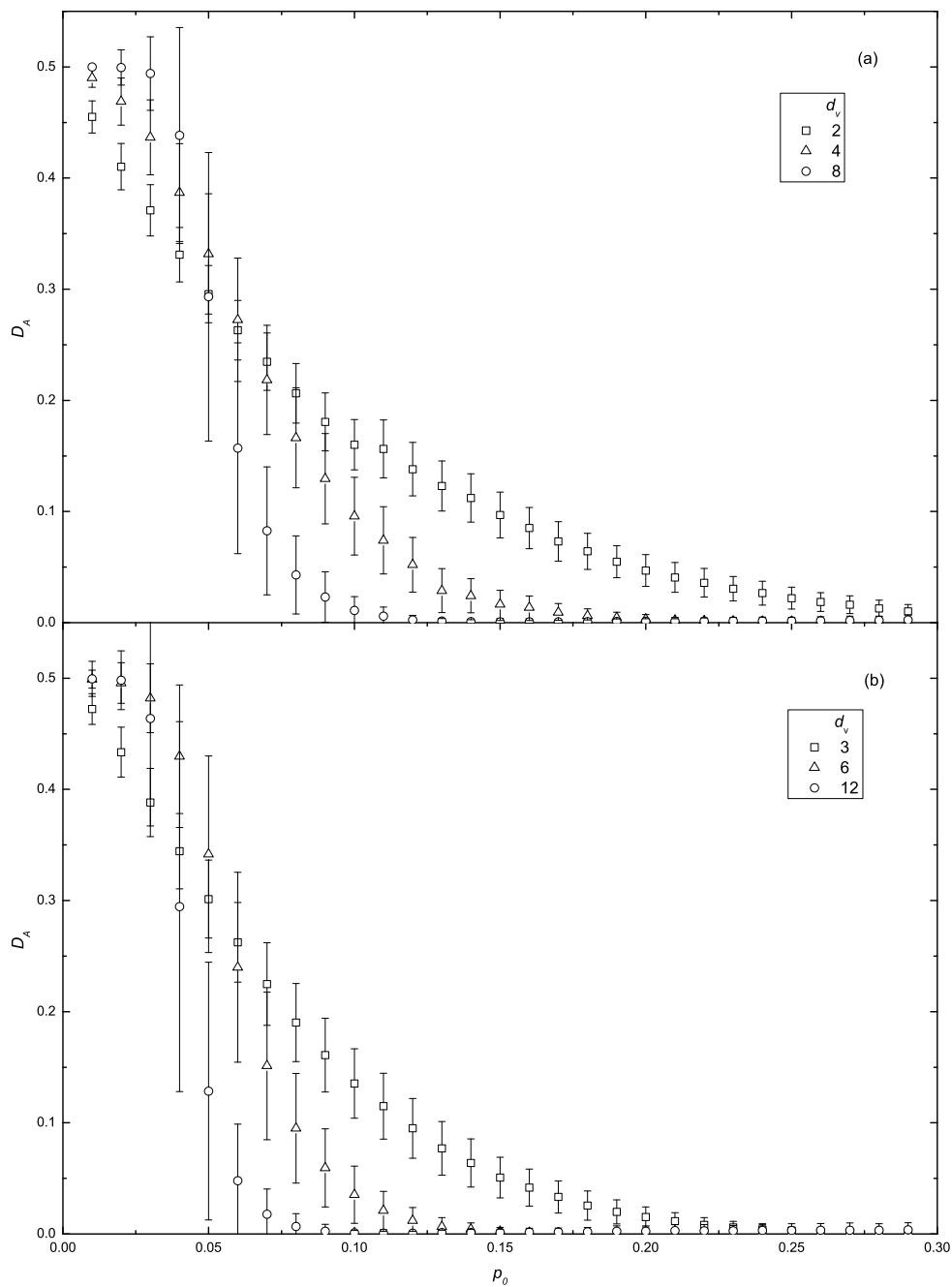}
 \caption{A plot of the yield $D_\textrm{A}$ of Implementation~A for
 different $d_v$'s versus the error probability $p_0$. The values of quantum
 rate and $n$ are fixed to $1/2$ and $960$ respectively.} 
\label{fig:ImA_r}
\end{figure*}

Figs.~\ref{fig:ImA_r}a and~\ref{fig:ImA_r}b show the yields $D_\textrm{A}$ using
$(d_v, d_c)$-regular QLDPC codes for different $d_v$ at a fixed quantum code
rate of $1 - d_v / d_c = 1/2$. As $d_v$ increases, the distance of the
QLDPC code $H[1]$ increases. That is why the $D_\textrm{A}$
against $p_0$ curve is strictly decreasing for $d_c \lesssim 5$, indicating
that the code $H[1]$ is only error detecting.  In contrast,
this curve is flat for very small $p_0$ whenever $d_c \gtrsim 6$, indicating
that the code $H[1]$ becomes error correcting.
Note further that in the latter case, the
yield $D_\textrm{A}$ drops rapidly when $p_0$ increases beyond the flat
region of the yield curve.  This is a consequence of
our error rejection mechanism. Recall that those EPR pairs which have the
low belief on the kind of error experienced will be thrown
away. Since $d_v$ is small and $H[1]$ is sparse,
statistical fluctuations may allow Alice and Bob to correctly identify a few
non-erroneous pairs via the belief propagation algorithm. And these
correctly identified pairs will be kept, making the error rate of the remaining
pairs lower than that of the original $n$ EPR pairs.
In other words, by increasing $d_v$ while keeping the ratio $d_v/d_c$ fixed,
it is harder to identify
of these non-erroneous pairs as the Tanner graph associated with $H$ becomes
more connected. Furthermore, this increase in the connectivity of the
Tanner graph implies that backward propagation of quantum errors as a result
of error syndrome measurement is more serious. This is also a contributing
factors to the low yield as $d_v$ increases by keeping $d_v / d_c$ fixed.

\begin{figure*}
\centering
\includegraphics*[bb = 20 25 490 662, scale = 0.8]{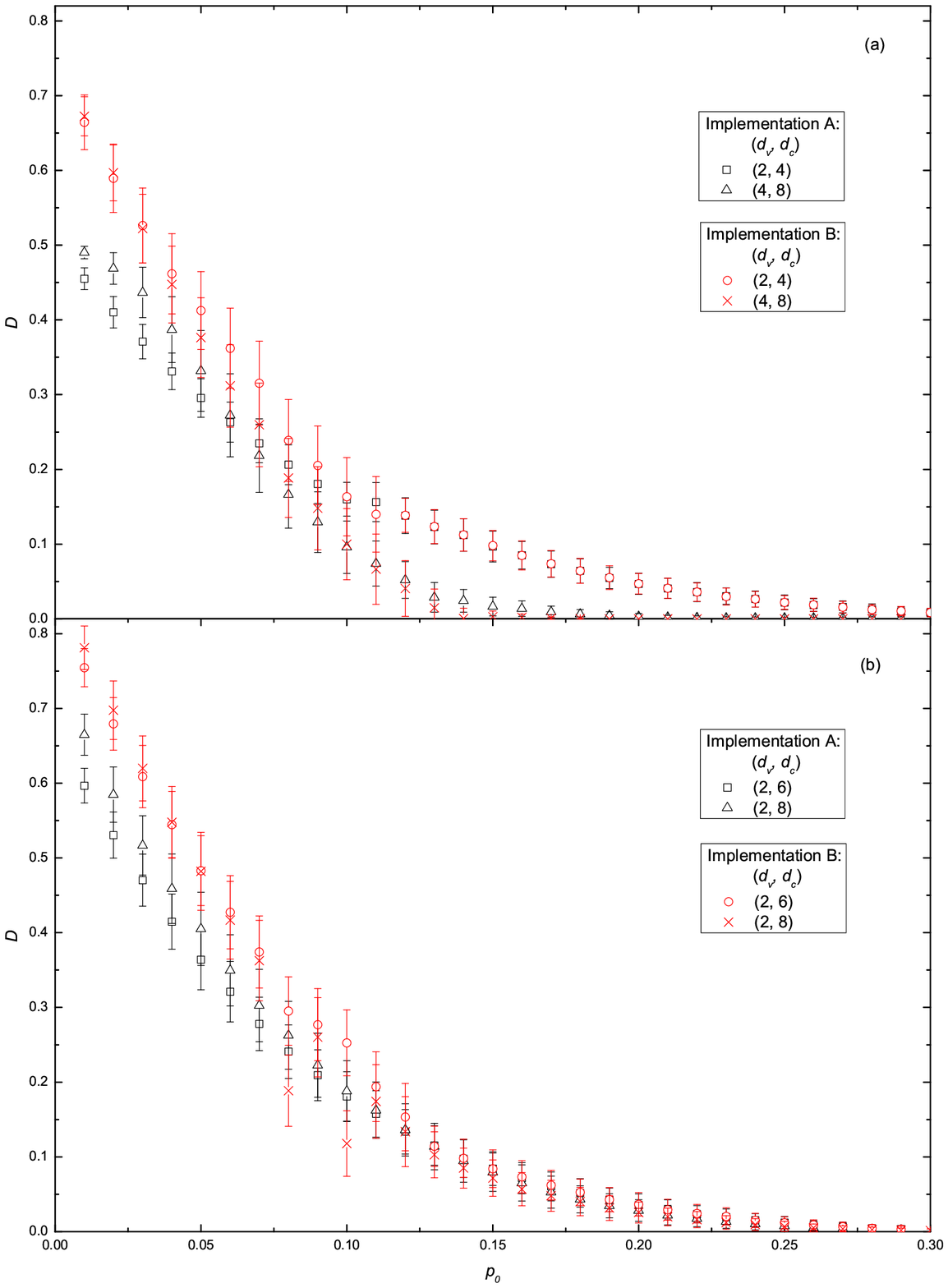}
 \caption{Comparison of the yield $D$ of Implementation~A and~B. (a) The
 codes are fixed at quantum rate $1/2$. (b) The $d_v$ for codes 
 are fixed at $2$. }
\label{fig:ImAB}
\end{figure*}

Fig.~\ref{fig:ImAB} shows the comparison of yields of our Implementation~A and~B.
In line with our expectation, for the same set of parameters $d_v$, $d_c$ and $n$, Implementation~B
outperforms Implementation~A for small $p_0$ where their performances converge as $p_0$ increases.
The adaptive nature of Implementation~B allows Alice and Bob to pick a quantum code that is
sufficiently powerful to combat the channel noise on the one hand and is sufficiently high rate to give
a good yield on the other hand. This demonstrates the power of adaptation using 
${\mathfrak P}_\textrm{BP}$. Nevertheless, adaption of this kind cannot
improve the capacity of ${\mathfrak P}_\textrm{BP}$ when $p_0$ is large.
In fact, Fig.~\ref{fig:ImAB} shows that Implementations~A and~B can handle
the same maximum error rate provided that $H[1] = H$. (More precisely,
$D_B(p_0) \geq D_A(p_0)$, but $D_B(p_0) = 0$ if and only if $D_A(p_0) = 0$.)
This finding is not completely surprising. When $p_0$ is sufficiently high,
it is likely that tentative decoding for $H[1], H[2], \ldots, 
h[\ell_\textrm{max} - 1]$ cannot be found. Thus, in this regime, Implementation~B
is reduced to Implementation~A.

\begin{figure*}
\centering
\includegraphics*[bb = 10 10 461 395]{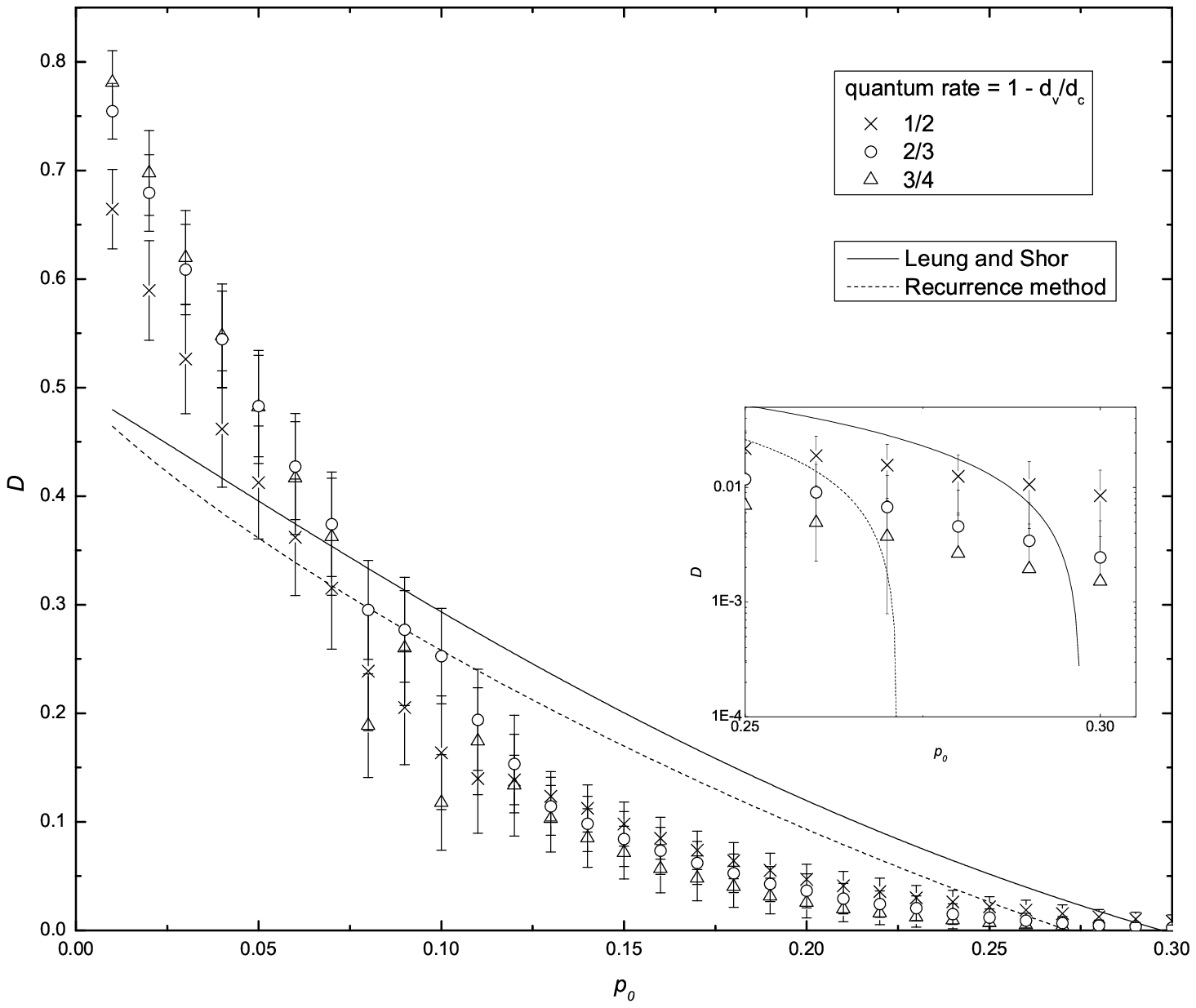}
 \caption{A plot of the yield $D_\textrm{B}$ of Implementation~B using
 $(d_v,d_c)$-regular QLDPC codes for different quantum rates against
 the error probability $p_0$. The values of $d_v$ and $n$ are fixed to be
 $2$ and $960$, respectively.}
\label{fig:ImB_dv2}
\end{figure*}

In Fig.~\ref{fig:ImB_dv2}, we compare the performances of Implementation~B
using $(2, d_c)$-regular QLDPC codes with the recurrence method~\cite{BDSW96a}
and Leung and Shor's protocol~\cite{LS07a}. Using $(2, 8)$-regular QLDPC 
codes, the yield of Implementation~B is higher than that of the recurrence method as 
well as the the Leung and Shor's protocol whenever $p_0 < 0.07$ because at 
such small $p_0$, there is no need to perform so heavy different parity checks. Similar
to the case of Implementation~A, we find that the maximum tolerable
rate for Implementation~B using $(2, 4)$-regular QLDPC codes is higher
than both the recurrence and the Leung and Shor protocols provided that
the channel error rate $p_0 > 0.28$.

\section{Conclusions And Outlook} 
\label{Sec:Conc}
In conclusion, we have introduced an adaptive two-way entanglement distillation
protocol ${\mathfrak P}_\textrm{BP}$ using QLDPC codes and belief propagation
decoding algorithm. In particular, we demonstrate the power of using QLDPC codes
and/or adaptation in EDP2 protocol. Moreover, we find that the yield of our scheme
${\mathfrak P}_\textrm{BP}$ using an ensemble of $(2,4)$-regular QLDPC codes is
higher than that of Leung and Shor~\cite{LS07a,LS07b} for handling depolarization error
whenever the error probability $p_0$ is greater than $0.28$ or less than or
equal to $0.07$.  In fact, using
this choice of QLDPC codes, our scheme ${\mathfrak P}_\textrm{BP}$ can tolerate
depolarization errors up to at least a quantum error rate of 30\%.

The high yield together with the reasonably high error tolerance capability of
our scheme are due to several reasons.  First, the scheme is adaptive in the
sense that each parity check may be used for error rejection or error
correction depending on the measurement results.  In this way our scheme
becomes an effectively one-way QECC-based scheme when the quantum error rate is
low.  And on other hand, it becomes a EDP2 based scheme when the quantum error
rate is high.

Note that as long as Alice and Bob find that a certain qubit has experienced a
certain quantum error with high probability, they can apply the necessary
error correction operation to recover a high fidelity EPR pair.  What causes
the trouble is that sometimes Alice and Bob are unable to correctly determine
or have little confidence on the kind of errors has occurred in their shared
qubits.  This leads us to the second reason why our scheme
${\mathfrak P}_\textrm{BP}$ is so efficient.  This is due to the fact that the
belief propagation decoding algorithm is able to efficiently find out the
entropy of the kind of quantum errors believed to be experienced by which of
the qubits are higher than the entropy threshold $h_\textrm{th}[\ell]$.
Consequently, Alice and Bob may throw this kind of qubits away.  The belief
propagation approach is a Bayesian approach.  It takes into account Alice and
Bob's initial belief on the channel and the information obtained from the
error syndrome measurements in a transparent way in computing the final belief
of the errors occurred.  By choosing a sufficiently long codeword size $n$, it
is highly probable that the error rate of a few variable nodes that are
connected to a check node in the corresponding Tanner graph is significantly
lower than the system average.  The belief propagation decoding algorithm can
help Alice and Bob to identify these variable nodes.  By selectively keeping
this kind of variable nodes (that is, these qubits) for entanglement
distillation, our EDP2 protocol ${\mathfrak P}_\textrm{BP}$ is able to
tolerate a reasonably high quantum error rate.

Finally, the last reason behind the good performance of
${\mathfrak P}_\textrm{BP}$ lies in the use of QLDPC codes.  It ensures that
the average error correcting capability of its punctured code is still
acceptable.

A number of followup researches along this line have to be done.  For instance,
the effect of the choice of the entropy thresholds $h_{\textrm{th}}[\ell]$ on the
yield $D_\textrm{BP}$ should be studied.  More importantly, the choice of
QLDPC codes with sparse parity check matrices $H[\ell]$'s in a multi-level
(that is, $\ell_\textrm{max} > 1$) setting requires thorough investigations in
order to understand the capability and the tradeoff between the yield
and the maximum error tolerable rate of our protocol. In particular,
we believe that the maximum error tolerable rate for 
${\mathfrak P}_\textrm{BP}$ can be pushed up further by adaptively
concatenating QLDPC codes.

\section*{Acknowledgment}
Valuable discussions with H.-K. Lo and C.-H. F. Fung are gratefully acknowledged.  This work is supported by the
RGC grant number HKU701004 of the HKSAR government. We would like to thank the
Computer Center of HKU for their helpful support in providing the use of the
HPCPOWER System for most of the simulations reported in this paper.

\bibliographystyle{IEEEtran}
\bibliography{qc40.1}
\end{document}